\documentclass[runningheads]{llncs}

\usepackage{cite}
\usepackage{caption}
\usepackage[T1]{fontenc}
\usepackage{graphicx}
\usepackage{booktabs}
\usepackage[colorlinks,linkcolor=blue,anchorcolor=blue,citecolor=blue,urlcolor=blue]{hyperref}
\usepackage{makecell}
\usepackage{multirow}
\usepackage{enumerate}
\usepackage[doipre={DOI:}]{uri}
\usepackage{siunitx}
\usepackage[ruled,vlined,linesnumbered]{algorithm2e}
\usepackage{overpic}
\usepackage{multirow}
\usepackage{amsfonts}
\usepackage{amsmath}
\usepackage{subfig}
\usepackage{graphicx}
\usepackage{grffile}
\usepackage{makecell}
\usepackage{subcaption}
\usepackage{pifont}
\usepackage{bbding}
\newcommand{\cmark}{\Checkmark} 
\newcommand{\xmark}{\XSolidBrush} 
\newcommand{\qmark}{\textbf{under-annotated}}

\usepackage{color}

\SetKw{KwBreak}{break}

\begin{document}

\title{mmRAG: A Modular Benchmark for Retrieval-Augmented Generation over Text, Tables, and Knowledge Graphs}
\titlerunning{mmRAG: A Modular Benchmark for Retrieval-Augmented Generation}

\author{
    Chuan~Xu
    \orcidID{0009-0002-3410-9664} \and
    Qiaosheng~Chen
    \orcidID{0009-0002-0610-7725} \and
    Yutong~Feng
    \orcidID{0009-0004-1553-1485} \and
    Gong~Cheng
    \orcidID{0000-0003-3539-7776}
}
\authorrunning{C. Xu et al.}

\institute{
State Key Lab for Novel Software Technology, Nanjing University, Nanjing, China 
\email{\{221240097, qschen, ytfeng\}@smail.nju.edu.cn, gcheng@nju.edu.cn}
}

\maketitle

\begin{abstract}
Retrieval-Augmented Generation (RAG) has emerged as a powerful paradigm for enhancing the capabilities of large language models. However, existing RAG evaluation predominantly focuses on text retrieval and relies on opaque, end-to-end assessments of generated outputs. To address these limitations, we introduce mmRAG, a modular benchmark designed for evaluating multi-modal RAG systems. Our benchmark integrates queries from six diverse question-answering datasets spanning text, tables, and knowledge graphs, which we uniformly convert into retrievable documents. To enable direct, granular evaluation of individual RAG components---such as the accuracy of retrieval and query routing---beyond end-to-end generation quality, we follow standard information retrieval procedures to annotate document relevance and derive dataset relevance. We establish baseline performance by evaluating a wide range of RAG implementations on mmRAG.
\end{abstract}

\section{Introduction}
\label{sec:introduction}

Retrieval‑Augmented Generation (RAG) has significantly advanced open-domain question answering (ODQA) by incorporating an external retriever into a large language model (LLM) to perform up-to-date and more reliable text-to-text generation~\cite{RAGMEETLLM, ragsurvey}. RAG applications increasingly demand reasoning over heterogeneous knowledge sources, such as knowledge graphs (KGs)~\cite{gretriever, chatkbqa}. However, existing RAG benchmarks remain largely \emph{single-modal}~\cite{benchmarkingLLM, ragbench} and evaluate RAG systems with \emph{end‑to‑end metrics} that obscure whether failures arise in generation or retrieval~\cite{BERGEN, compmix}. Moreover, none of them supports the evaluation of \emph{query routing}, which allows RAG systems to identify and retrieve from a particular source, reducing the retrieval cost. These limitations prevent comprehensive diagnosis and optimization of individual components of an RAG system.

\paragraph{Our Work:}
To address the above limitations of existing RAG benchmarks, we introduce \textbf{mmRAG}, a \emph{multi-modal} and \emph{modular} benchmark designed to evaluate the main components of RAG beyond generation, \emph{including query routing and retrieval}. We integrate six diverse QA datasets that span text, tables, and KGs into a unified corpus of retrievable documents. We provide cross-dataset annotations of relevance labels to evaluate retrieval accuracy and derive dataset-level relevance labels to evaluate query routing accuracy. Our benchmark comprises \num{5124}~queries, 3.2~million chunks from \num{90998}~documents, and \num{88751}~annotated query-chunk pairs, offering a unique testbed for modular evaluation of multi-modal RAG. Its novel features are summarized as follows.
\begin{itemize}
  \item \textbf{Unified multi‑modal corpus:} Chunks are sourced and converted from a hybrid of text, tables, and KGs.
  \item \textbf{Cross-dataset relevance annotation:} Queries are annotated with relevant chunks from all datasets to directly assess retrieval accuracy.
  \item \textbf{Whole-process modular evaluation:} Annotations are provided separately for query routing, retrieval, and generation to support the direct evaluation of these individual RAG components.
\end{itemize}

\paragraph{Availability:}
We clarify the mandatory availability of our resource as follows.
\begin{itemize}
    \item The mmRAG benchmark is published at Hugging Face with a DOI.\footnote{\url{https://doi.org/10.57967/hf/5475}}
    \item The mmRAG benchmark has a canonical citation~\cite{mmragds}.
    \item The mmRAG benchmark is open under the Apache License 2.0.
\end{itemize}

\paragraph{Outline:}
The remainder of this paper is organized as follows. Section~\ref{sec:related-work} surveys related benchmarks. Section~\ref{sec:approach} details our benchmark construction. Section~\ref{sec:experiments-ret} and Section~\ref{sec:routerexp} report evaluation results. Section~\ref{sec:conclusion} concludes the paper with future directions and the Resource Availability Statement.
\section{Related Work}
\label{sec:related-work}


\begin{table}[t]
\centering
\caption{Single-modal ODQA and RAG benchmarks.}
\label{tab:single}
\small
\resizebox{\linewidth}{!}{
\begin{tabular}{lccccccc}
\toprule
\textbf{Benchmark} & \multicolumn{4}{c}{\textbf{Modality}} & \multicolumn{3}{c}{\textbf{Available Labels}} \\
\cmidrule(lr){2-5} \cmidrule(lr){6-8}
 & \textbf{Text} & \textbf{Table} & \textbf{Image} & \textbf{KG} & \textbf{Generation} & \textbf{Retrieval} & \textbf{Query Routing} \\
\midrule
WebQuestions~\cite{webqsp}      & \xmark & \xmark & \xmark & \cmark & \cmark & \xmark & \xmark \\
OK-VQA~\cite{OKVQA}        & \xmark & \xmark & \cmark & \xmark & \cmark & \xmark & \xmark \\
NQ~\cite{NQ} & \cmark & \xmark & \xmark & \xmark & \cmark & \qmark & \xmark \\
HotpotQA~\cite{hotpotqa}          & \cmark & \xmark & \xmark & \xmark & \cmark & \qmark & \xmark \\
KILT~\cite{KILT}              & \cmark & \xmark & \xmark & \xmark & \cmark & \qmark & \xmark \\
RAGBench~\cite{ragbench}          & \cmark & \xmark & \xmark & \xmark & \cmark & \cmark & \xmark \\
\bottomrule
\end{tabular}
}
\end{table}

The early ODQA and RAG benchmarks laid important groundwork, but remain confined to single modality and often under‑annotate relevance signals. Table~\ref{tab:single} summarizes representative single‑modal RAG benchmarks, their supported modalities, and the presence of labels for generation, retrieval, and query routing. These six datasets can be divided into three tiers based on the type of annotation. First, generation‑only benchmarks such as WebQuestions~\cite{webqsp} and OK-VQA~\cite{OKVQA} provide generation labels, with neither retrieval nor query routing annotations. Next, Natural Questions (NQ)~\cite{NQ}, HotpotQA~\cite{hotpotqa}, and KILT~\cite{KILT} augment the generation labels with annotations of document relevance, but only for one or a few pertinent documents heuristically selected rather than systematically examined across the entire corpus, thus referred to as \emph{under-annotated}. They are insufficient to evaluate the accuracy of the retrieval. Finally, RAGBench~\cite{ragbench} provides both the generation and the sufficiently annotated retrieval labels, yet it still lacks annotations to evaluate query routing.

\begin{table}[t]
\centering
\caption{Multi-modal ODQA and RAG benchmarks.}
\label{tab:multi}
\small
\resizebox{\linewidth}{!}{
\begin{tabular}{lccccccc}
\toprule
\textbf{Benchmark} & \multicolumn{4}{c}{\textbf{Modality}} & \multicolumn{3}{c}{\textbf{Available Labels}} \\
\cmidrule(lr){2-5} \cmidrule(lr){6-8}
 & \textbf{Text} & \textbf{Table} & \textbf{Image} & \textbf{KG} 
 & \textbf{Generation} & \textbf{Retrieval} & \textbf{Query Routing} \\
\midrule
HybridQA~\cite{hybridqa}      & \cmark & \cmark & \xmark & \xmark & \cmark & \qmark & \xmark \\
DEXTER~\cite{dexter}        & \cmark & \cmark & \xmark & \xmark & \cmark & \qmark & \xmark \\
OTT-QA~\cite{OTT}        & \cmark & \cmark & \xmark & \xmark & \cmark & \cmark & \xmark \\
KVQA~\cite{KVQA}          & \xmark & \xmark & \cmark & \cmark & \cmark & \qmark & \xmark \\
FVQA~\cite{fvqa}          & \xmark & \xmark & \cmark & \cmark & \cmark & \qmark & \xmark \\
CompMix~\cite{compmix}       & \cmark & \cmark & \xmark & \cmark & \cmark & \xmark & \xmark \\
MultiModalQA~\cite{multimodalqa}  & \cmark & \cmark & \cmark & \xmark & \cmark & \qmark & \xmark \\
\midrule
mmRAG (ours) & \cmark & \cmark & \xmark & \cmark & \cmark & \cmark & \cmark \\
\bottomrule
\end{tabular}
}
\end{table}

Building on these single‑modal foundations, recent RAG benchmarks introduce heterogeneous data formats but still lack comprehensive support for modular evaluation. HybridQA\cite{hybridqa} and DEXTER~\cite{dexter} combine text with tables for multi‐hop reasoning. OTT‑QA\cite{OTT} further incorporates relevance labels for retrieval evaulation. KVQA~\cite{KVQA} and FVQA~\cite{fvqa} combine images and KGs.  CompMix~\cite{compmix} and MultiModalQA~\cite{multimodalqa} expand to tri‐modal corpora. As Table~\ref{tab:multi} shows, most multi‑modal RAG benchmarks still under‑annotate retrieval labels and none of them provides annotations to evaluate query routing.

\emph{Compared with existing RAG benchmarks, our mmRAG not only covers three modalities in one unified suite but also supports modular evaluation: sufficiently annotated relevance labels are provided to evaluate retrieval and query routing.} This combined feature characterizes the uniqueness of our benchmark.



\section{Construction of mmRAG}
\label{sec:approach}

\subsection{Overview}
\label{sec:ap-overview}

\begin{figure}[!t]
    \centering
    \includegraphics[width=1\columnwidth]{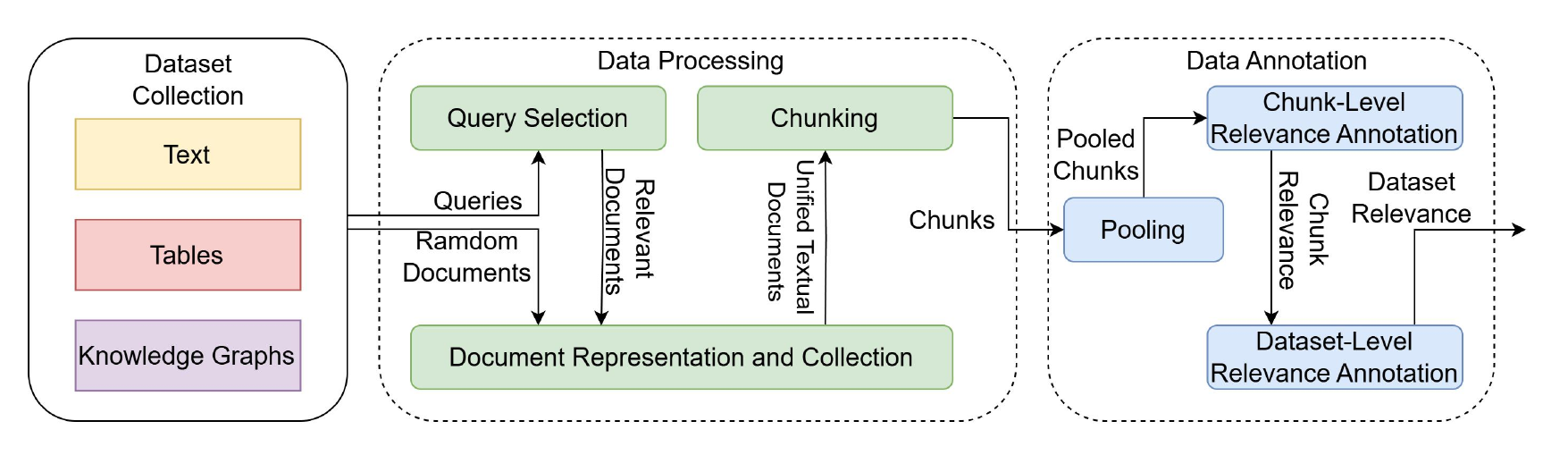}
    \caption{Construction of mmRAG.}
    \label{fig:flowchart}
\end{figure}

As shown in Figure~\ref{fig:flowchart}, the construction of our mmRAG benchmark follows three phases. \textbf{Dataset Collection} selects six diverse QA datasets that span text, tables, and KGs, incorporating real-world user queries and complex reasoning tasks to provide a foundation for evaluating RAG systems in real-world scenarios that require accurate information retrieval (IR) and multi-modal reasoning~(Section~\ref{sec:ap-dataset-collection}). \textbf{Data Processing} employs three core techniques. First, we select representative queries from these QA datasets using clustering techniques to capture various information needs and reduce redundancy. Next, we build a collection of textual documents converted from different data formats, including documents relevant to the queries and random documents as noise to simulate real-world retrieval scenarios. Finally, we segment documents into small chunks to improve compatibility with various retrieval architectures~(Section~\ref{sec:ap-data-processing}). \textbf{Data Annotation} follows standard IR protocols, combining methods based on pooling with LLM-based automatic annotation, to provide sufficient relevance labels for retrieval evaluation. Based on chunk-level annotations, we derive dataset-level relevance labels to enable the evaluation of query routing~(Section~\ref{sec:ap-data-annotation}).

\begin{table}[h]
    \centering
    \caption{Characteristics of datasets included in mmRAG.}
    \label{tab:dataset_characteristics}
    \small
    \begin{tabular}{lcccc}
        \toprule
        \textbf{Dataset} & \textbf{Modality} & \textbf{Domain} & \textbf{Query Source} & \textbf{Reasoning Task} \\
        \midrule
        NQ~\cite{NQ}             & Text         & Open-domain & Query log   & Single-hop        \\
        TriviaQA~\cite{triviaqa} & Text         & Open-domain & Crowdsourcing  & Single- \& multi-hop    \\
        OTT~\cite{OTT}        & Table+Text   & Open-domain & Query log         & Multi-hop         \\
        TAT~\cite{tat}        & Table+Text   & Financial   & Query log    & Numerical         \\
        CWQ~\cite{cwq}           & KG           & Open-domain & Crowdsourcing            & Multi-hop            \\
        WebQSP~\cite{webqsp}     & KG           & Open-domain & Query log           & Single- \& multi-hop            \\
        \bottomrule
    \end{tabular}
\end{table}

\subsection{Dataset Collection}
\label{sec:ap-dataset-collection}

To evaluate RAG systems, we collect six diverse QA datasets. As shown in Table~\ref{tab:dataset_characteristics}, they collectively contribute the following characteristics to our benchmark. These features underpin the rationality of our dataset collection, offering a foundation for assessing the multi-faceted capabilities of RAG systems.

\paragraph{Multiple Data Modalities:}
We include datasets over text (NQ~\cite{NQ}, TriviaQA~\cite{triviaqa}), tables (OTT~\cite{OTT}, TAT~\cite{tat}), and KGs (CWQ~\cite{cwq}, WebQSP~\cite{webqsp}), ensuring varied data formats to test different retrieval capabilities.

\paragraph{Real User Queries:}
We select datasets that contain natural queries derived from real-world interactions. For example, NQ captures real search queries and TriviaQA provides Trivia-style questions. By emphasizing such real user interactions, our benchmark better reflects practical retrieval scenarios.


\paragraph{Diverse Reasoning Tasks:}
For a thorough evaluation of RAG systems, it is essential to test their ability to handle different tasks. Our benchmark encompasses a wide range of reasoning tasks, including multi-hop QA and numerical reasoning.

\subsection{Data Processing}
\label{sec:ap-data-processing}

To build an unbiased multi-source, multi-modal benchmark, our data processing is organized into three phases: (1) Query Selection, (2) Document Representation and Collection, and (3) Chunking.

\begin{table}[t]
    \centering
    \caption{Statistics of mmRAG.}
    \label{tab:dataset_stats}
    \small
    \begin{tabular}{p{5cm}rrr}
        \toprule
        \textbf{Source Dataset}   & \textbf{Queries} & \textbf{Documents} & \quad\textbf{Chunks} \\
        \midrule
        NQ           & 990     & 16,000   & 1,448,163  \\
        TriviaQA     & 931     & 15,000   & 470,488    \\
        OTT          & 953     & 27,798   & 52,541     \\
        TAT          & 589     & 2,201    & 3,242      \\
        KG (CWQ + WebQSP)          & 834 + 827 & 29,999   & 1,227,114  \\
        \midrule
        Total & 5,124& 90,998 & 3,201,548 \\
        \bottomrule
    \end{tabular}
\end{table}

\subsubsection{Query Selection}

From each dataset, we sample a subset of queries to collectively form our query set. To achieve both representativeness and diversity, we extract all queries with a non‑empty answer from each dataset, embed them into a semantic space, and group them into $\num{1000}$~clusters using K‑means as in~\cite{mair}. To further refine the selection, we adopt an LLM-based filtering mechanism inspired by~\cite{DBLP:journals/corr/abs-2410-10594}. This filtering step aims to eliminate queries that are overly context-dependent, thereby ensuring that our selected queries remain meaningful beyond any specific context to fit the multi-source nature of our benchmark. Within each cluster, the first query retained by the filter is accepted as the representative of the cluster. Table~\ref{tab:dataset_stats} presents the final number of queries accepted from each dataset. The total number~5,124 has excluded 125~queries that---latter in data annotation---are not associated with any relevant chunk.

\subsubsection{Document Representation and Collection}

We transform the corpora of all original datasets into a unified document representation.

For KG-based datasets, observe that both CWQ and WebQSP are based on the Freebase KG. For each query, we identify all the binding values of the variables in the corresponding SPARQL query. For each binding value that is an entity, we construct a document to represent an one-hop subgraph centered around this entity in the KG, including both incoming and outgoing edges. We select and verbalize edges (i.e., triples) in the following order: (1) triples in the gold-standard SPARQL query results to ensure that the document contains the answer; (2) edges labeled with \texttt{rdf:type}; (3) edges linked to literal values; and (4) a random sample of the remaining edges---at most \num{10000}---to add as many relationships to other entities as possible. This document serves as a focused representation of the entity and its immediate relationships within the KG. Following this process, we construct \num{15359} documents from Freebase. They encapsulate the immediate knowledge of entities that are relevant to the selected queries. To add noise documents, we further construct \num{14640} documents from the neighborhood of these entities. They are not directly involved in any query, but are connected to some relevant entities, forming hard negatives. In summary, from KG-based datasets we construct \num{29999} documents.

For the datasets based on text and tables, we directly adopt the textual documents provided by the original datasets. For each query, we collect all the relevant documents in the original dataset. We further augment the document collection with documents randomly sampled from the datasets such that the total number of documents from text-based datasets (NQ and TriviaQA), $16,000+15,000=31,000$, and the number of documents from table-based datasets (OTT and TAT), $27,798+2,201=29,999$, are comparable to the number mentioned above of documents from KG-based datasets, as summarized in Table~\ref{tab:dataset_stats}, to form a balanced distribution across different modalities.

The total number of documents is~90,998.

\subsubsection{Chunking}

To ensure compatibility with the limited input capacity of dense retrievers, all documents are segmented into fixed-length chunks using a token-based splitter. Specifically, we employ the token splitter provided by the LangChain framework,\footnote{\url{https://python.langchain.com/api_reference/text_splitters/base/langchain_text_splitters.base.TokenTextSplitter.html}} which partitions each document into non-overlapping chunks of 512~tokens. Each chunk is assigned a unique identifier, allowing efficient indexing, retrieval, and evaluation at the chunk level. As Table~\ref{tab:dataset_stats} presents, we obtain a total of \num{3201548} chunks, representing a large corpus to retrieve.

\subsection{Data Annotation}
\label{sec:ap-data-annotation}

For each query, we annotate its relevant chunks using the standard IR pooling method and then derive relevance labels at the dataset level.

\subsubsection{Pooling}
It is impractical to annotate the relevance of 3.2~million chunks to 5,124~queries. As a common practice in IR, pooling significantly reduces the number of chunks required to be annotated and ensures that the vast majority of relevant chunks are annotated, assuming the remaining ones irrelevant.

Specifically, each query is processed with two popular yet complementary retrievers: BM25 and BGE-large-en-v1.5~\cite{bgeembedding}, which respectively capture exact lexical matches and semantic similarities. The chunk pool for each query consists of up to 19~top-ranked chunks retrieved by each retriever, including the following.
\begin{itemize}
    \item Globally top-10 chunks.
    \item Top-3 chunks from the relevant document in the original dataset. This ensures that the most pertinent chunks---those the query was originally meant to hit---are always present in the pool.
    \item Top-1 chunk from each dataset. This is important to capture all possible dataset-level relevance, since we will later derive dataset-level relevance labels from chunk-level annotations.
\end{itemize}

The final pool for each query is created by combining these three subsets from the two retrievers. For each query, an average of 17.31~chunks are pooled to be annotated. Unpooled chunks are assumed to be irrelevant to the query.

\subsubsection{Chunk-Level Relevance Annotation}

Given a query \( q \) and a pooled chunk~\( c \), we annotate with a three-level graded relevance label \( L_{q, c} \in \{0, 1, 2\} \), representing irrelevant, partially relevant, and highly relevant (i.e., providing useful context), respectively~\cite{procis}, through an ensemble scheme involving two primary annotators and one tiebreaker. The primary annotators are two powerful and cost-effective LLMs: DeepSeek-V3\cite{deepseekai2024deepseekv3technicalreport} and GLM-4-Plus\footnote{\url{https://open.bigmodel.cn/dev/api/normal-model/glm-4}} (or Claude-3.5-Sonnet\footnote{\url{https://www.anthropic.com/news/claude-3-5-sonnet}} when GLM-4-Plus occasionally does not respond). If the two primary annotators disagree, GPT-4o\footnote{\url{https://openai.com/index/hello-gpt-4o/}} will be used as a tiebreaker to determine the final label. Specifically, if GPT-4o agrees with either primary annotator, that label is taken. If all three annotators disagree with each other, we will take their average value, that is, $L_{q, c}=1$. This protocol ensures that each query-chunk pair is evaluated by at least two strong models, with GPT-4o used only when necessary due to its relatively high cost.

\paragraph{Annotator Agreement:}
We analyze the agreement between the annotators. Among all the \num{90846} query-chunk pairs, the two primary annotators produce the same relevance label on \num{77363} pairs (85\%), representing a significant level of agreement. All three annotators disagree with each other only on \num{479} pairs (0.53\%). \emph{These numbers suggest the high quality of our annotations.}\\

Table~\ref{tab:relevance_distribution} presents the distribution of the final labels. For 125~queries, all their pooled chunks are irrelevant. These queries are excluded from our benchmark.

\begin{table}[t]
  \centering
  \caption{Distribution of chunk-level relevance labels per query in each dataset.}
  \label{tab:relevance_distribution}
\small
  \begin{tabular}{lrrrr}
    \toprule
    \textbf{Dataset} & \multicolumn{4}{c}{\textbf{Relevance Label}}  \\
    \cmidrule(lr){2-5}
    & \textbf{0} & \textbf{1} & \textbf{2} & \textbf{Total}  \\
    \midrule
    NQ       & 10.91 & 3.26 & 4.52 & 18.70\\
    TriviaQA & 9.50  & 1.68 & 4.74 & 15.92\\
    OTT      & 11.83 & 3.83 & 0.94 & 16.61\\
    TAT      & 11.30 & 3.91 & 2.51 & 17.72\\
    CWQ      & 8.84  & 5.14 & 3.59 & 17.57\\
    WebQSP   & 7.66  & 4.19 & 5.74 & 17.60\\
    \midrule
    Overall  & 10.02 & 3.59 & 3.70 & 17.31\\
    \bottomrule
  \end{tabular}
\end{table}

\begin{table}[t]
  \centering
  \caption{Proportion of queries in each dataset having relevant chunks in other datasets.}
  \label{tab:label_distribution}
\small
  \begin{tabular}{lrrrrr}
    \toprule
    \textbf{Original Dataset} & \multicolumn{5}{c}{\textbf{Target Dataset}} \\
    \cmidrule(lr){2-6}
    & {NQ} & {TriviaQA} & {OTT} & {TAT} & {KG (CWQ + WebQSP)} \\
    \midrule
NQ       & 99.80\% & 49.80\%   & 14.55\% & 0.51\%  & 20.91\% \\
TriviaQA & 62.41\% & 93.34\%   & 18.05\% & 0.43\%  & 26.53\% \\
OTT      & 27.81\% & 20.78\%   & 97.06\% & 0.00\%  & 18.78\% \\
TAT      & 21.73\% & 15.62\%   & 4.41\%  & 100.00\%& 2.89\%  \\
CWQ      & 61.87\% & 55.28\%   & 26.86\% & 0.84\%  & 99.88\% \\
WebQSP   & 68.56\% & 64.93\%   & 30.35\% & 1.45\%  & 99.52\% \\
    \bottomrule
  \end{tabular}
\end{table}

\paragraph{Cross-Dataset Relevance:}
A query may have relevant chunks in both its original dataset and other datasets. Table~\ref{tab:label_distribution} shows the proportion of queries that have relevant chunks ($L_{q, c} \geq 1$) in other datasets.
Although it is trivial that diagonal entries are close to~100\%, we are interested in off-diagonal entries that represent cross-dataset relevance. We have two key observations. First, there is pronounced cross‑dataset relevance, as queries from one dataset frequently (up to~68.56\%) retrieve pertinent chunks from other datasets, \emph{showing that our effort to cross-dataset relevance annotation is essential to the measurement of retrieval accuracy, which is often missing in existing RAG benchmarks}. Second, TAT is rarely involved in cross-dataset relevance, which is not surprising because this dataset is for the financial domain, while the others are open-domain datasets.

\subsubsection{Dataset-Level Relevance Annotation}

Building upon chunk-level annotations, we introduce dataset-level relevance labels to provide a reference for query routing. This label quantifies the contribution of each dataset to answering a specific query, reflecting the alignment between the query and the dataset's content. To obtain this label, we aggregate the relevance signals from individual chunks. Specifically, given a query \( q \) and a dataset \( D \), we annotate the relevance of~$D$ to~$q$ with the following dataset-level relevance label:
\begin{equation}
    S_{q, D} = \sum_{d \in D} \max_{c \in d}{L_{q, c}} \,,
\end{equation}
\noindent where $c \in d$ means $c$~is a chunk split from document~$d$, and \( L_{q, c} \) denotes the chunk-level relevance label for \( c \) with respect to \( q \). The idea here is to aggregate the relevance of documents in~$D$, and for each document we only consider its most relevant chunk to avoid distorting the result by long documents.


\begin{table}[t]
  \centering
  \caption{Mean dataset-level relevance label ($S_{q, D}$) per query in each dataset.}
  \label{tab:mean_dataset_label}
\small
  \begin{tabular}{lrrrrr}
    \toprule
    \textbf{Original Dataset of~$q$} & \multicolumn{5}{c}{\textbf{Target Dataset~($D$)}} \\
    \cmidrule(lr){2-6}
    & {NQ} & {TriviaQA} & {OTT} & {TAT} & {KG (CWQ + WebQSP)} \\
    \midrule
NQ       & 4.28 & 1.20 & 0.28 & 0.01 & 0.34 \\
TriviaQA & 1.99 & 3.43 & 0.41 & 0.01 & 0.48 \\
OTT      & 0.48 & 0.32 & 4.09 & 0.00 & 0.24 \\
TAT      & 0.35 & 0.19 & 0.05 & 7.44 & 0.04 \\
CWQ      & 1.64 & 1.10 & 0.47 & 0.01 & 3.05 \\
WebQSP   & 2.36 & 1.75 & 0.63 & 0.02 & 3.60 \\
    \bottomrule
  \end{tabular}
\end{table}

Table~\ref{tab:mean_dataset_label} presents the distribution of the derived dataset-level relevance labels. In particular, many off-diagonal entries exceed~1, indicating that, on average, each query in these datasets finds at least one relevant chunk in a different dataset, highlighting the practicality of cross-dataset relevance. The overall distribution aligns with the chunk-level distribution in Table~\ref{tab:label_distribution}, with the highest values on the diagonal and between particular pairs of dataset such as WebQSP-NQ and TriviaQA-NQ, further supporting the need for query routing. \emph{With our dataset-level relevance labels, routing accuracy can be directly measured, which is not enabled by previous RAG benchmarks.}

\subsection{Data Splits}

For a fair comparison between different users of our mmRAG benchmark in the future, we provide an official split of our data into train/dev/test sets in a 60\%/15\%/25\% ratio. We split 5,124~queries by stratified sampling so that these three sets follow approximately the same distribution of datasets. There are 3,072~queries in the train set, 766~in the dev set, and 1,286~in the test set.
\section{Evaluation of Retrievers}
\label{sec:experiments-ret}

We can indirectly evaluate retrievers by using the original query answers and assessing generation quality, or directly measure retrieval accuracy based on the relevance labels provided by our mmRAG benchmark. In this section, we employ mmRAG to evaluate popular retrievers to establish a baseline for future research.

\subsection{Evaluation Setup}

\subsubsection{Retrievers}
We evaluate a diverse set of retrieval models, from classic lexical methods to modern neural models.

Classic retrievers include three widely used IR baselines:
\begin{itemize}
  \item \textbf{BM25}, a lexical ranking function known for its efficiency and robustness,
  \item \textbf{Contriever~\cite{contriever}}, a dense retriever trained with contrastive learning to produce semantically rich embeddings, and
  \item \textbf{DPR~\cite{dpr}}, a bi‐encoder trained on query-passage pairs.
\end{itemize}
We use public checkpoints of Contriever\footnote{\url{https://huggingface.co/facebook/contriever/tree/main}} and DPR\footnote{\url{https://huggingface.co/docs/transformers/model_doc/dpr}} without further fine-tuning them on mmRAG.

Modern retrievers include
\begin{itemize}
  \item \textbf{BGE~\cite{bgeembedding}}, i.e. \texttt{bge-large-en-v1.5}, a generative encoder selected for its leading results on the MTEB leaderboard,\footnote{\url{https://huggingface.co/spaces/mteb/leaderboard}}
  \item \textbf{GTE~\cite{gte1, gte2}}, i.e. \texttt{gte-large-en-v1.5}, a Transformer‐based generative encoder also ranked among the best on MTEB, and
  \item \textbf{Fine‐tuned BGE} and \textbf{Fine-tuned GTE}, both fine‐tuned on the train and valid sets of mmRAG for 1~epoch with hard negatives.
\end{itemize}

We also set up an \textbf{Oracle} retriever that always outputs an optimal ranking and achieves perfect retrieval accuracy. We use it as a reference when measuring the quality of downstream generation.

\subsubsection{Generators}
We combine the above retrievers with two popular LLMs of different sizes as generation models:
\begin{itemize}
  \item \textbf{GLM~\cite{glm}}, i.e. \texttt{glm-4-plus}, a large, online‐accessible LLM, and
  \item \textbf{Qwen~\cite{qwen2, qwen2.5}}, i.e. \texttt{Qwen-7B-Instruct}, a 7-billion-parameter, locally deployable LLM, representing a resource‐constrained setting. 
\end{itemize}
\noindent We prompt them with top-3 retrieved chunks to augment generation.


\subsubsection{Evaluation Metrics}

We measure retrieval accuracy and generation quality.
\begin{itemize}
    \item For retrieval accuracy, we use three standard IR metrics reported at cut-offs $k = 1, 3, 5$: Normalized Discounted Cumulative Gain (\textbf{NDCG@$k$}), Mean Average Precision (\textbf{MAP@$k$}), and \textbf{Hits@$k$} (i.e.,~the proportion of queries that have at least one relevant chunk in the top-$k$). For MAP and Hits which are based on binary relevance labels, we define relevant as $L_{q, c} \geq 1$.
    \item For \textbf{generation quality}, we use Exact Match (EM) for datasets with a single correct answer (TriviaQA, OTT, WebQSP) and use the F1 score for datasets with multiple correct answers (NQ, TAT, CWQ).
\end{itemize}
\noindent We report these metrics averaged over all queries in the test set of mmRAG.

\begin{table}[t]
  \centering
  \caption{Evaluation of retrievers (retrieval accuracy).}
  \label{tab:main_retrieval}
  \small
  \resizebox{\linewidth}{!}{
  \begin{tabular}{lccccccccc}
    \toprule
    \textbf{Retriever}            & \textbf{NDCG@1} & \textbf{MAP@1} & \textbf{Hits@1} & \textbf{NDCG@3} & \textbf{MAP@3} & \textbf{Hits@3} & \textbf{NDCG@5} & \textbf{MAP@5} & \textbf{Hits@5} \\
    \midrule
    BM25            & 0.531           & 0.102          & 0.612          & 0.525           & 0.241          & \underline{1.726} & \underline{0.534} & 0.345          & \underline{2.725} \\
    Contriever      & 0.216           & 0.043          & 0.245          & 0.201           & 0.087          & 0.611             & 0.195             & 0.109          & 0.880             \\
    DPR             & 0.121           & 0.020          & 0.138          & 0.114           & 0.040          & 0.358             & 0.110             & 0.050          & 0.513             \\
    BGE             & \textbf{0.617}  & \underline{0.114}          & \textbf{0.703} & \textbf{0.607}  & \textbf{0.273} & \textbf{1.971}    & \textbf{0.618}    & \textbf{0.395} & \textbf{3.107}    \\
    GTE             & 0.452           & 0.089          & 0.500          & 0.416           & 0.186          & 1.251             & 0.398             & 0.235          & 1.767             \\
    Fine-tuned BGE  & \underline{0.591} & \textbf{0.116} & \underline{0.664} & \underline{0.542} & \underline{0.269} & 1.669           & 0.523             & \underline{0.355} & 2.397             \\
    Fine-tuned GTE  & 0.526           & 0.105          & 0.584          & 0.487           & 0.226          & 1.481             & 0.467             & 0.286          & 2.082             \\
    \bottomrule
  \end{tabular}
  }
\end{table}

\begin{table}[!h]
  \centering
  \caption{Evaluation of retrievers (generation quality).}
  \label{tab:main_generation}
  \small
  \begin{tabular}{lccccccc}
    \toprule
    \textbf{Retriever} & \textbf{NQ} & \textbf{TriviaQA} & \textbf{OTT} & \textbf{TAT} & \textbf{CWQ} & \textbf{WebQSP} & \textbf{Avg.} \\
    \midrule
    \emph{Used with GLM}\\
    No retrieval     & 0.2782 & \textbf{0.6239} & 0.0625 & 0.0212 & 0.2511 & 0.2415 & 0.2464 \\
    BM25             & 0.2379 & 0.5299 & 0.1375 & \underline{0.1757} & 0.4162 & 0.2560 & 0.2922 \\
    Contriever       & 0.2500 & \underline{0.5855} & 0.0833 & 0.1149 & 0.2465 & 0.2754 & 0.2593 \\
    DPR              & 0.2258 & 0.4915 & 0.0583 & 0.0541 & 0.2143 & 0.1932 & 0.2062 \\
    BGE              & 0.2903 & 0.5342 & 0.1167 & 0.1419 & 0.3007 & 0.2705 & 0.2757 \\
    GTE              & \underline{0.3065} & 0.5385 & 0.1333 & 0.1284 & 0.3704 & 0.2947 & 0.2953 \\
    Fine-tuned BGE   & \textbf{0.3387} & 0.5769 & \underline{0.1458} & 0.1338 & \underline{0.4653} & \underline{0.4106} & \underline{0.3452} \\
    Fine-tuned GTE   & \underline{0.3065} & 0.5641 & \textbf{0.1750} & \textbf{0.1811} & \textbf{0.4956} & \textbf{0.4203} & \textbf{0.3571} \\
    Oracle           & {0.3548} & 0.5769 & {0.2458} & {0.2723} & {0.5920} & {0.4444} & {0.4145} \\
    \midrule
    \emph{Used with Qwen}\\
    No retrieval      & 0.1008 & 0.4060 & 0.0417 & 0.0358 & 0.1861 & 0.0870 & 0.1429 \\
    BM25              & 0.1613 & 0.4231 & 0.0542 & 0.1622 & 0.3339 & 0.1643 & 0.2165 \\
    Contriever        & 0.2056 & 0.3846 & 0.0417 & 0.1588 & 0.1706 & 0.1498 & 0.1852 \\
    DPR               & 0.1734 & 0.2821 & 0.0250 & 0.0405 & 0.1128 & 0.1208 & 0.1258 \\
    BGE               & 0.2661 & 0.4145 & 0.0708 & 0.1351 & 0.2300 & 0.2174 & 0.2223 \\
    GTE               & 0.2782 & \textbf{0.4774} & 0.0625 & 0.1622 & 0.2545 & 0.3043 & 0.2482 \\
    Fine-tuned BGE    & \textbf{0.3306} & \underline{0.4744} & \underline{0.0833} & \underline{0.1811} & \underline{0.3857} & \underline{0.4058} & \textbf{0.3102} \\
    Fine-tuned GTE    & \underline{0.2863} & 0.4359 & \textbf{0.1000} & \textbf{0.1946} & \textbf{0.4225} & \textbf{0.4203} & \underline{0.3099} \\
    Oracle            & {0.3185} & {0.5085} & {0.1625} & {0.2095} & {0.5353} & {0.4976} & {0.3720} \\
    \bottomrule
  \end{tabular}
\end{table}

\subsection{Main Evaluation Results}


\subsubsection{Retrieval Accuracy}
As shown in Table~\ref{tab:main_retrieval}, BGE exhibits the strongest performance with NDCG@1 of 0.617 and Hits@1 of 0.703, largely outperforming the other retrievers. GTE is in the middle range with NDCG@1 of 0.452, while its fine-tuned version shows a notable improvement of 0.074 in NDCG@1 and 0.084 in Hits@1. Classic dense retrievers such as Contriever and DPR appear less competitive in this experiment.

\subsubsection{Generation Quality}
In Table~\ref{tab:main_generation}, RAG generally outperforms direct generation without retrieval, underscoring the importance of high-quality retrieval. Used with GLM, BM25 leads to an average score of 0.2922, and better results are obtained with fine-tuned BGE and GTE, reaching~0.3452 and~0.3571, respectively. However, there is a gap between these retrievers and the Oracle retriever which achieves~0.4145, suggesting room for future studies. With Qwen, the absolute generation quality becomes lower, but the relative results remain similar.\\

When comparing the generation quality in Table~\ref{tab:main_generation} with the retrieval accuracy in Table~\ref{tab:main_retrieval}, the two metrics generally exhibit a positive correlation, with a few exceptions. For example, BM25 outperforms GTE in retrieval accuracy, while GTE leads to better generation quality. \emph{It indicates that direct and indirect evaluation of the retrieval in RAG present a degree of complementarity.}

\begin{table}[t]
  \centering
  \caption{Evaluation of retrievers (generation quality) over all chunks versus over dataset-specific chunks.}
  \label{tab:retrieval_full_vs_isolated}
  \small
  \resizebox{\textwidth}{!}{
  \begin{tabular}{lcccccccccccc}
    \toprule
    \textbf{Retriever} 
    & \multicolumn{6}{c}{\textbf{Over All Chunks}} 
    & \multicolumn{6}{c}{\textbf{Over Dataset-Specific Chunks}} \\
    \cmidrule(lr){2-7} \cmidrule(lr){8-13}
    & NQ & TriviaQA & OTT & TAT & CWQ & WebQSP
    & NQ & TriviaQA & OTT & TAT & CWQ & WebQSP \\
    \midrule
    \emph{Used with GLM}\\
    BM25   & 0.2379 & 0.5299 & 0.1375 & 0.1757 & 0.4162 & 0.2560 
           & 0.2661 & 0.5171 & 0.1583 & 0.1892 & 0.5210 & 0.3865 \\
    BGE    & 0.2903 & 0.5342 & 0.1167 & 0.1419 & 0.3007 & 0.2705 
           & 0.3105 & 0.5641 & 0.1500 & 0.1676 & 0.3835 & 0.3720 \\
    GTE    & 0.3065 & 0.5385 & 0.1333 & 0.1284 & 0.3704 & 0.2947 
           & 0.3266 & 0.5513 & 0.1375 & 0.1541 & 0.4169 & 0.3816 \\
    Oracle & 0.3548 & 0.5769 & 0.2458 & 0.2723 & 0.5920 & 0.4444 
           & 0.3548 & 0.5684 & 0.2167 & 0.2284 & 0.6435 & 0.5411 \\
    \midrule
    \emph{Used with Qwen}\\
    BM25   & 0.1613 & 0.4231 & 0.0542 & 0.1622 & 0.3339 & 0.1643 
           & 0.1815 & 0.4060 & 0.0667 & 0.1689 & 0.3901 & 0.3285 \\
    BGE    & 0.2661 & 0.4145 & 0.0708 & 0.1351 & 0.2300 & 0.2174 
           & 0.2944 & 0.4060 & 0.0833 & 0.1473 & 0.2932 & 0.3333 \\
    GTE    & 0.2782 & 0.4774 & 0.0625 & 0.1622 & 0.2545 & 0.3043 
           & 0.3185 & 0.4402 & 0.0750 & 0.1811 & 0.3171 & 0.3865 \\
    Oracle & 0.3185 & 0.5085 & 0.1625 & 0.2095 & 0.5353 & 0.4976 
           & 0.3427 & 0.4701 & 0.1458 & 0.2128 & 0.5370 & 0.5604 \\
    \bottomrule
  \end{tabular}
  }
\end{table}

\subsection{Evaluation of Dataset-Specific Retrieval}  

Our mmRAG benchmark integrates six QA datasets. In this experiment, we explore how generation quality varies when we restrict retrieval to the chunks from the original dataset of each query, i.e., only retrieving \emph{dataset-specific chunks}. Due to resource constraints, here we only experiment with a subset of the best-performing retrievers in previous experiments, including BM25, BGE, GTE, and the Oracle retriever. The results are compared in Table~\ref{tab:retrieval_full_vs_isolated}.


From dataset-specific chunks to all chunks, the generation quality with the Oracle retriever increases on TriviaQA and OTT. It means that additional documents from other datasets---possibly in a different modality---can provide contexts that more helpfully augment generation than the original documents. \emph{This observation encourages future research on cross-modal or multi-modal RAG which is currently still under-explored.}

However, such quality increases are rarely seen on other retrievers. Indeed, with BM25, BGE, and GTE, the generation quality generally drops considerably when expanding the scope of the retrieval from dataset-specific chunks to all chunks. For example, with GLM, BM25 drops from~0.5210 over CWQ chunks to~0.4162 over all chunks, BGE drops from~0.3720 over WebQSP chunks to~0.2705 over all chunks, and GTE declines from~0.3266 over NQ chunks to~0.3065 over all chunks. \emph{This performance decline demonstrates that the integration of multiple datasets of different modalities in mmRAG raises new challenges to RAG.} \emph{This performance difference also underscores the importance of query routing to RAG systems}, which we will evaluate with mmRAG in the next section.

\section{Evaluation of Query Routers}
\label{sec:routerexp}

We can indirectly evaluate query routers by using the original query answers and assessing generation quality, or directly measure routing accuracy based on the dataset-level relevance labels provided by our mmRAG benchmark. In this section, we employ mmRAG to evaluate several baseline query routers.

\subsection{Evaluation Setup}

\subsubsection{Retrievers and Generators}
Following previous experiments, we use three retrievers: \textbf{BM25}, \textbf{BGE}, \textbf{GTE}, which achieve relatively high retrieval accuracy in previous experiments. We use \textbf{Qwen} as our LLM generator because its generation quality is more sensitive to retrieval accuracy than GLM, so it can better reflect the influence of query routing. We prompt it with top-3 retrieved chunks to augment generation.

\subsubsection{Query Routers}
Query routing has not been extensively studied in the literature. We evaluate two existing routing methods.
\begin{itemize}
    \item \textbf{Semantic router} is inspired by an existing implementation.\footnote{\url{https://github.com/aurelio-labs/semantic-router}} We use BGE to encode both the query and the description of each of the five datasets, NQ, TriviaQA, OTT, TAT, and KG~(CWQ + WebQSP). Descriptions are collected from the dataset homepages. We calculate the cosine similarity between two encoding vectors as the routing score used in ranking datasets.
    \item \textbf{LLM router} is inspired by~\cite{llmrouter}. We prompt GLM to rank the five datasets in terms of their likelihood of containing the answer. Similar routing strategies are also used in LlamaIndex\footnote{\url{https://docs.llamaindex.ai/en/stable/module_guides/querying/router/}} and LangChain.\footnote{\url{https://python.langchain.com.cn/docs/modules/chains/foundational/router}}
\end{itemize}

Further, we set up an \textbf{Oracle router} that always outputs an optimal ranking of the datasets and achieves perfect routing accuracy. We use it as a reference when measuring the quality of downstream generation.

\subsubsection{Evaluation Metrics}
We measure routing accuracy and generation quality.
\begin{itemize}
    \item Similarly to previous experiments, we measure routing accuracy by \textbf{NDCG@$k$}, \textbf{MAP@$k$}, and \textbf{Hits@$k$} at cut-offs $k = 1, 2, 3, 4, 5$ which refers to the number of top-ranked datasets. For MAP and Hits which are based on binary relevance labels, we define relevant as $S_{q, D} \geq 1$.
    \item For \textbf{generation quality}, following previous experiments, we measure the EM or F1 score, depending on the dataset, at cut-offs $k = 1, 2, 3, 4, 5$. We configure the retrievers to only retrieve dataset-specific chunks to augment generation, i.e.,~those from the top-$k$ datasets.
\end{itemize}
\noindent We report these metrics averaged over all queries in the test set of mmRAG.

\begin{figure}[t]
    \centering
    \includegraphics[width=1\columnwidth]{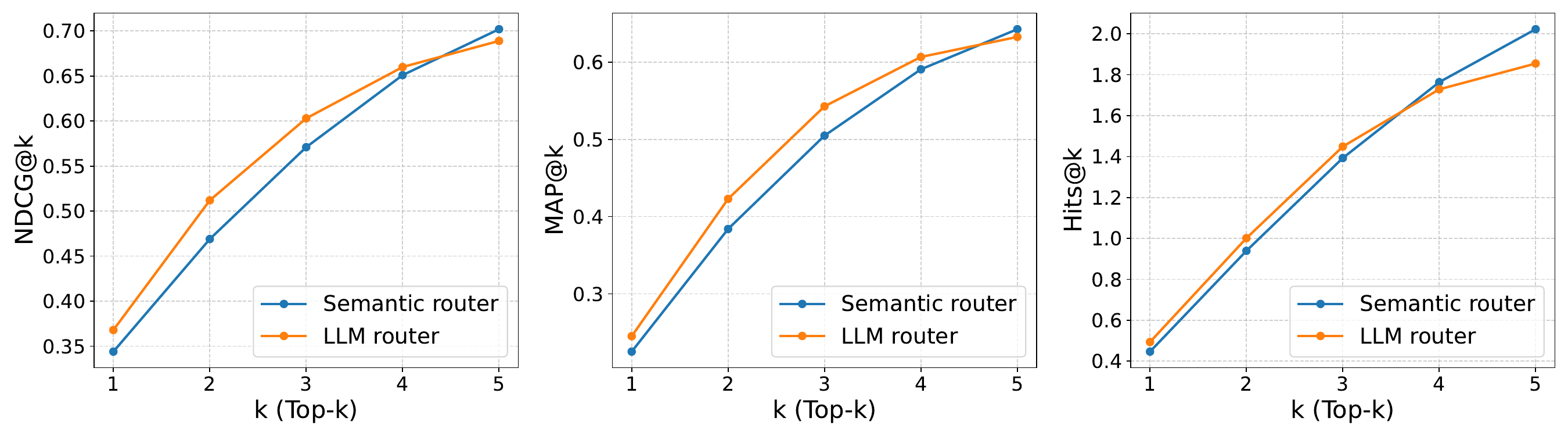}
    \caption{Evaluation of query routers (routing accuracy).}
    \label{fig:score_router}
\end{figure}

\begin{figure}[t]
    \centering
    \begin{minipage}[t]{0.3\textwidth}
        \centering
        \includegraphics[width=\textwidth]{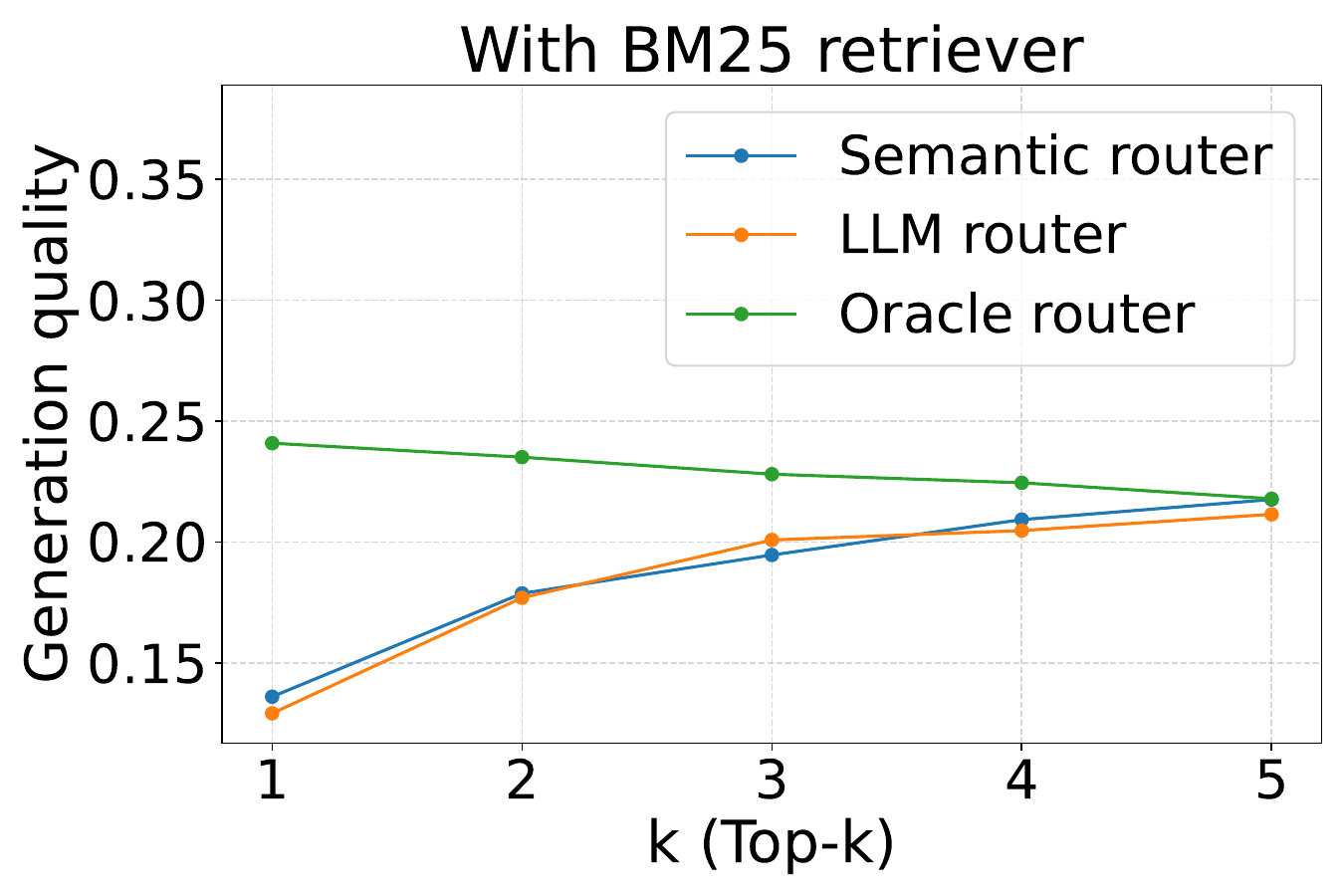}
    \end{minipage}
    \begin{minipage}[t]{0.3\textwidth}
        \centering
        \includegraphics[width=\textwidth]{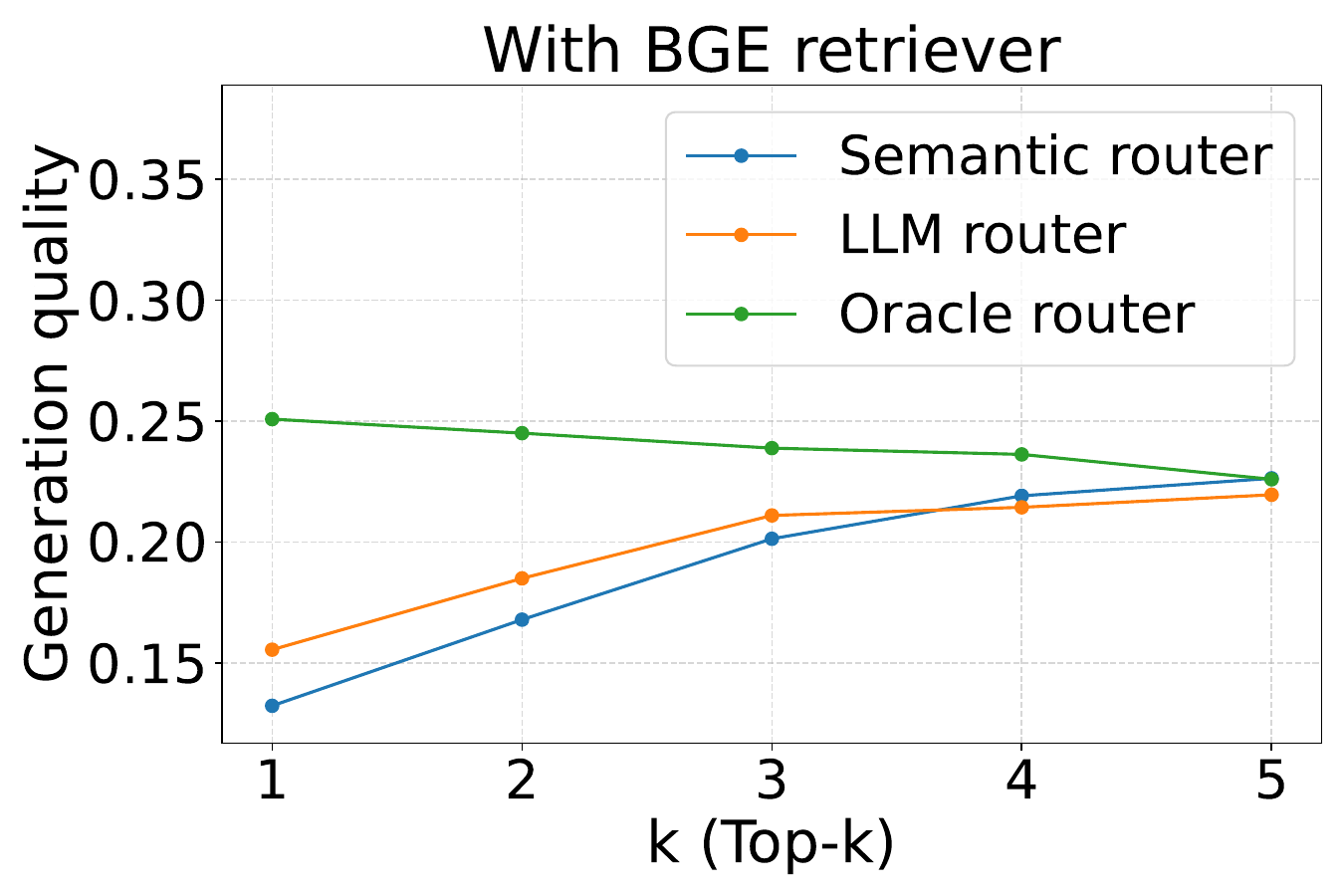}
    \end{minipage}
    \begin{minipage}[t]{0.3\textwidth}
        \centering
        \includegraphics[width=\textwidth]{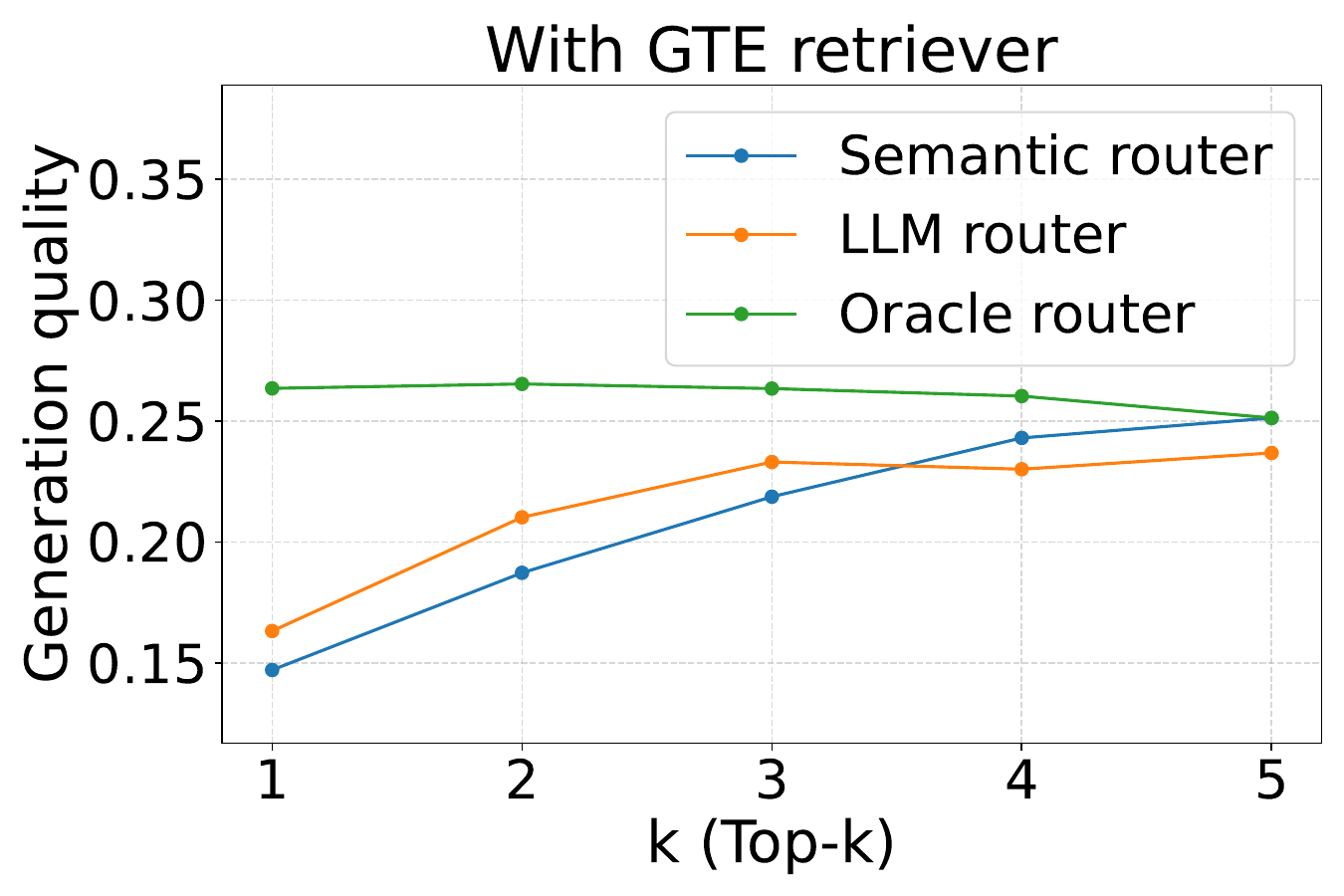}
    \end{minipage}
    \caption{Evaluation of query routers (generation quality).}
    \label{fig:gen_router}
\end{figure}

\subsection{Evaluation Results}

\subsubsection{Routing Accuracy}
As shown in Figure~\ref{fig:score_router}, in all three metrics, NDCG@$k$, MAP@$k$, and Hits@$k$, the LLM router consistently outperforms the semantic router at small values of~$k$ which represents the main application scenario of query routing where the query is sent to a small number of top-ranked datasets.

\subsubsection{Generation Quality}
As illustrated in Figure~\ref{fig:gen_router}, with a BM25 retriever, the LLM router and the semantic router lead to comparable generation quality. However, with the BGE and GTE retrievers, the LLM router helps to achieve higher generation quality than the semantic retriever at small values of~$k$. Compared with the Oracle router, the gaps are noticeable, suggesting significant room for future studies on query routing.\\


When comparing the generation quality in Figure~\ref{fig:gen_router} with the routing accuracy in Figure~\ref{fig:score_router}, the general trends of these two metrics appear similar. \emph{It indicates that, with the  dataset-level relevance labels provided by mmRAG, direct measurement of routing accuracy can serve as a cost-effective alternative to indirect evaluation with generation quality which is computationally expensive.}

\section{Conclusion}
\label{sec:conclusion}

In this work, we present a significant advancement in RAG benchmarking by shifting from single-modal, end-to-end evaluation to a multi-modal, modular framework. In contrast to existing RAG benchmarks that focus on text retrieval or only evaluate end-to-end generation quality, our mmRAG integrates text, tables, and KGs with high-quality relevance annotations to directly evaluate retrieval accuracy. Furthermore, mmRAG is among the first to support the evaluation of query routing in RAG systems by providing relevance labels at the dataset level. Together with the original gold-standard query answers, these multi-stage annotations enable direct, modular evaluation of individual RAG components including query routing, retrieval, and generation, offering a way to comprehensively analyze the performance of RAG systems. The multi-modal nature of mmRAG, covering KGs and other data formats that are commonly used on the Web and in knowledge-centric applications, will also encourage a wider adoption of Semantic Web technologies.

To foster community adoption, we publish mmRAG and our codebase with detailed documentation and tutorials. We anticipate that the multi-modal and modular characteristics of mmRAG will benefit a wide range of research in cross-domain and structured data QA~\cite{surveyOnComplexFactualQA} and inspire future innovations in RAG. Beyond its primary use as a benchmark, mmRAG also offers valuable signals for related tasks. Its dataset-level annotations can support query router training and facilitate the quality assessment of metadata generated for QA datasets.

\paragraph{Limitations and Future Work:}
While mmRAG demonstrates notable strengths, several limitations remain.
First, it currently supports only three data modalities---text, tables, and KGs---lacking visual modalities such as images. Future extensions will explore the incorporation of richer multi-modal content to support a more comprehensive RAG evaluation.
Second, our LLM-based annotation process is computationally expensive and time-consuming, which limits its scalability. Improving the efficiency of this procedure is a key direction for our future work.
Third, mmRAG does not offer domain-specific data splits, which would enable a more fine-grained evaluation and analysis of query routing strategies. Its future releases may include such partitions to facilitate domain-targeted studies. Furthermore, expanding the datasets to cover specialized domains such as medicine and law would support the development of vertical-domain benchmarks, extending the evaluation of RAG systems in knowledge-intensive applications.
In general, we plan to extend mmRAG toward both broader modality coverage and deeper domain specialization, strengthening its value as a modular benchmark for vertical and multi-modal RAG systems.

\paragraph*{Resource Availability Statement:}
The mmRAG benchmark data is available from Hugging Face~\cite{mmragds}. Source code related to mmRAG is available from GitHub at \url{https://github.com/nju-websoft/mmRAG}. All resources are available under the Apache License 2.0.



\clearpage
\appendix

\bibliographystyle{splncs04}
\bibliography{main}

\end{document}